%% file: paper.tex
\def\useextern{}
\documentclass[runningheads]{llncs}

\usepackage{ifluatex}

\usepackage{comment}

\usepackage{graphicx}

\usepackage{float}

\usepackage{hyperref}
\hypersetup{
	unicode,
	colorlinks = true,
	allcolors = black,
	urlcolor = blue,
	pdfborder = 0 0 0,
}

\usepackage{etoolbox}
\ifdef{\genextern}{%
\usepackage[newfloat,draft=false,chapter,finalizecache]{minted}
}{%
\ifdef{\useextern}{%
\usepackage[newfloat,draft=false,chapter,frozencache]{minted}
}{%
\usepackage[newfloat,draft=false,chapter]{minted}
}}%

\usepackage[newfloat,draft=false,chapter]{minted}
\setminted{fontsize=\footnotesize}
\newmintinline{c}{style=bw}
\newmintinline{text}{style=bw}
\usepackage{multirow}

\usepackage{subcaption}

\usepackage{longtable}

\usepackage[nameinlink,capitalize]{cleveref}
\crefname{sublisting}{Listing}{Listings}
\Crefname{sublisting}{Listing}{Listings}

\usepackage{minibox}

\ifdefined\useextern
\usepackage{tikzexternal}
\else
\usepackage{tikz}

\ifdefined\genextern
\usetikzlibrary{external}
\fi

\usetikzlibrary{calc}
\usetikzlibrary{decorations}
\usetikzlibrary{decorations.pathreplacing}
\usetikzlibrary{decorations.text}
\usetikzlibrary{decorations.pathmorphing}
\usetikzlibrary{decorations.markings}
\usetikzlibrary{fit}
\usetikzlibrary{arrows}    
\usetikzlibrary{shapes}
\usetikzlibrary{shadows}
\usetikzlibrary{automata}
\usetikzlibrary{positioning}
\usetikzlibrary{petri}
\usetikzlibrary{chains}
\usetikzlibrary{fadings}
\usetikzlibrary{matrix}
\usetikzlibrary{patterns}
\usetikzlibrary{mindmap}
\usetikzlibrary{graphs}			
\usetikzlibrary{graphdrawing}	
\usetikzlibrary{quotes}			
\usetikzlibrary{babel}          
\usetikzlibrary{hobby}
\usetikzlibrary{arrows.meta}
\usetikzlibrary{backgrounds}
\usetikzlibrary{bending}
\usetikzlibrary{external}

\usegdlibrary{trees}			
\usegdlibrary{layered}			
\usegdlibrary{force}			
\usegdlibrary{circular}			
\usegdlibrary{routing}			

\usepackage{pgfplots}
\pgfplotsset{compat=1.12}

\makeatletter
\def\myfoldheight{0.5}

\def\myshapepath{
	\pgfextract@process\northwest{
		\southwest\pgf@xa=\pgf@x
		\northeast
		\pgf@x=\pgf@xa
	}
	
	\pgfextract@process\southeast{
		\southwest\pgf@ya=\pgf@y
		\northeast
		\pgf@y=\pgf@ya
	}
	
	\pgfextract@process\northfold{
		\pgfpointdiff{\southwest}{\northeast}
		\northeast
		\advance\pgf@x-\myfoldheight\pgf@y
	}
	
	\pgfextract@process\eastfold{
		\pgfpointdiff{\southwest}{\northeast}
		\northeast
		\advance\pgf@y-\myfoldheight\pgf@y
	}
	
	\pgfextract@process\fold{
		\northfold\pgf@xa=\pgf@x
		\eastfold
		\pgf@x=\pgf@xa
	}
	
	\pgfpathmoveto{\southwest}
	\pgfpathlineto{\northwest}
	\pgfpathlineto{\northfold}
	\pgfpathlineto{\eastfold}
	\pgfpathlineto{\southeast}
	\pgfpathclose
}

\def\myshapeanchorborder#1#2{
	\pgftransformreset 
	\pgf@relevantforpicturesizefalse 
	\pgfintersectionofpaths{
		\myshapepath
	}{
		\pgfpathmoveto{
			\pgfpointadd{
				\pgfpointdiff{\southwest}{\northeast}\pgf@xc=\pgf@x \advance\pgf@xc by \pgf@y 
				\pgfpointscale{
					\pgf@xc
				}{
					\pgfpointnormalised{
						#2
					}
				}
			} {
				#1
			}
		}
		\pgfpathlineto{#1}
	}
	\pgfpointintersectionsolution{1}
}

\newdimen\myshapedimenx
\newdimen\myshapedimeny

\pgfdeclareshape{file}{
	\inheritsavedanchors[from=rectangle]
	\inheritanchor[from=rectangle]{center}
	\inheritanchor[from=rectangle]{mid}
	\inheritanchor[from=rectangle]{base}
	
	\inheritanchor[from=rectangle]{west}
	\inheritanchor[from=rectangle]{east}
	\inheritanchor[from=rectangle]{north}
	\inheritanchor[from=rectangle]{south}
	
	\inheritanchor[from=rectangle]{south west}
	\inheritanchor[from=rectangle]{south east}
	\inheritanchor[from=rectangle]{north west}
	
	\inheritanchor[from=rectangle]{mid west}
	\inheritanchor[from=rectangle]{mid east}
	\inheritanchor[from=rectangle]{base west}
	\inheritanchor[from=rectangle]{base east}
	
	\backgroundpath{
		\myshapepath
	}
	
	\foregroundpath{
		\pgfpathmoveto{\northfold}
		\pgfpathlineto{\fold}
		\pgfpathlineto{\eastfold}
	}
	
	\anchorborder{%
		\pgf@xb=\pgf@x
		\pgf@yb=\pgf@y%
		\southwest%
		\pgf@xa=\pgf@x
		\pgf@ya=\pgf@y%
		\northeast%
		\advance\pgf@x by-\pgf@xa%
		\advance\pgf@y by-\pgf@ya%
		\pgf@xc=.5\pgf@x
		\pgf@yc=.5\pgf@y%
		\advance\pgf@xa by\pgf@xc
		\advance\pgf@ya by\pgf@yc%
		\edef\pgf@marshal{%
			\noexpand\pgfpointborderrectangle
			{\noexpand\pgfqpoint{\the\pgf@xb}{\the\pgf@yb}}
			{\noexpand\pgfqpoint{\the\pgf@xc}{\the\pgf@yc}}%
		}%
		\pgf@process{\pgf@marshal}%
		\advance\pgf@x by\pgf@xa%
		\advance\pgf@y by\pgf@ya%
	}
	
}

\makeatother

\usepackage[textsize=miniscule,obeyDraft]{todonotes}
\makeatletter
\newcommand\miniscule{\@setfontsize\miniscule{4}{5}}
\makeatother

\fi

\let\svthefootnote\thefootnote
\newcommand\blankfootnote[1]{%
	\let\thefootnote\relax\footnotetext{#1}%
	\let\thefootnote\svthefootnote%
}

\usepackage{pgfmath}

\makeatletter
\newsavebox{\my@scalepar@TempBox}

\makeatother

\begin{document}
\title{Loop Optimization Framework}
%
%
\author{Michael Kruse, Hal Finkel}
%
%
\institute{Argonne Leadership Computing Facility,\\
Argonne National Laboratory, Argonne IL 60439, USA\\
\email{mkruse@anl.gov, hfinkel@anl.gov}}
\maketitle              
\begin{abstract}
The LLVM compiler framework supports a selection of loop transformations such as vectorization, distribution and unrolling.  
Each transformation is carried-out by specialized passes that have been developed independently.

In this paper we propose an integrated approach to loop optimizations:
A single dedicated pass that mutates a Loop Structure DAG.
Each transformation can make use of a common infrastructure such as dependency analysis, transformation preconditions, etc.

\end{abstract}

\input{text}

%
%
\bibliographystyle{splncs04}
\bibliography{bibliography}

\end{document}

%% file: text.tex

\definecolor{darkbrown}{rgb}{0.4, 0.26, 0.13}
\definecolor{dartmouthgreen}{rgb}{0.05, 0.5, 0.06}
\definecolor{crimsonglory}{rgb}{0.75, 0.0, 0.2}
\definecolor{chromeyellow}{rgb}{1.0, 0.65, 0.0}
\definecolor{amethyst}{rgb}{0.6, 0.4, 0.8}
\definecolor{darkgoldenrod}{rgb}{0.72, 0.53, 0.04}
\definecolor{darkmagenta}{rgb}{0.55, 0.0, 0.55}

\section{Shortcomings of Clang/LLVM's Pass Pipeline}

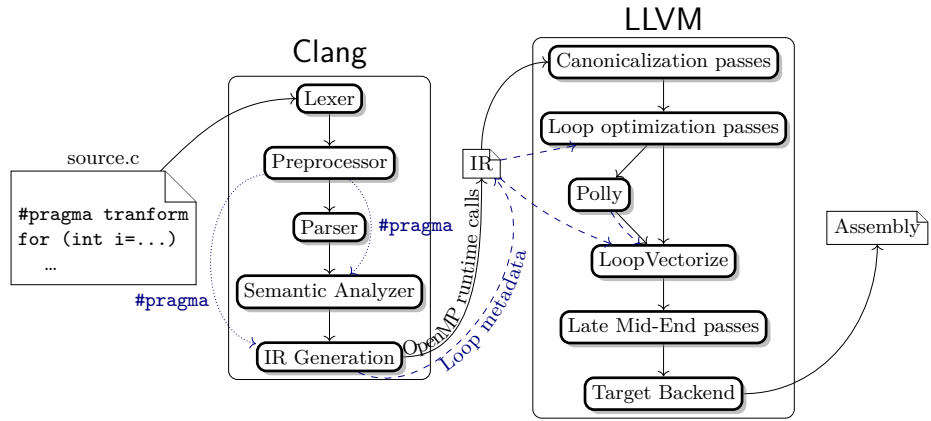
\begin{figure}[t]
	\resizebox{\linewidth}{!}{%
		\begin{tikzpicture}%
		\tikzset{tight/.style={inner sep=0pt,outer sep=0pt,minimum size=0pt}}
		\tikzset{node/.style={draw,fill=white,line width=1.2pt,rounded corners,drop shadow}}
		\tikzset{supernode/.style={subgraph text none,draw,rounded corners}}
		\tikzset{edge/.style={->}}
		\graph[layered layout,edges={edge,rounded corners},level sep=5mm,sibling sep=10mm]{
			c[as={\minibox{\\\texttt{\#pragma tranform}\\\texttt{for (int i=...)}\\\hspace*{4mm}\dots}},grow=right,draw,shape=file,fill=white,label={source.c}];
			ir[as={IR},shape=file,draw,fill=white,nudge=(up:10mm)];
			asm[as={Assembly},shape=file,draw,fill=white];
			
			clang [subgraph text none,label={[font=\Large\sffamily]above:Clang}] // [sibling sep=2mm,grow=down,layered layout] {
				lexer [as={Lexer},node];
				parser [as={Parser},node,grow=down];
				preprocessor [as={Preprocessor},node];
				sema [as={Semantic Analyzer},node];
				codegen [as={IR Generation},node];
				lexer->preprocessor->parser->sema->codegen;
			};
			llvm [subgraph text none,label={[font=\Large\sffamily]above:LLVM}] // [sibling sep=2mm,grow=down,layered layout] {
				canonicalization [as={Canonicalization passes},node];
				loopopts [as={Loop optimization passes},node];
				polly [as={Polly},node];
				vectorization [as={LoopVectorize},node];
				latepasses [as={Late Mid-End passes},node];
				backend [as={Target Backend},node];
				canonicalization->loopopts->vectorization->latepasses->backend;
				loopopts->polly->vectorization;
			};
			c->[in=180]lexer;
			codegen->[out=0,in=-90,postaction={decorate,decoration={text along path,text={OpenMP runtime calls},raise=0.3ex}}]ir;
			ir->[out=90,in=180]canonicalization;
			backend->[out=0,in=-90]asm;
		};
		\begin{pgfonlayer}{background}
		\node[tight,fit={(clang)},supernode]{};
		\node[tight,fit={(llvm)},supernode]{};
		\end{pgfonlayer}
		\path (preprocessor) edge[edge,densely dotted,bend left=50,draw=blue!50!black] node[midway,right,font=\ttfamily,blue!50!black] {\#pragma} (sema);
		\path (preprocessor) edge[edge,densely dotted,bend right=80,draw=blue!50!black] node[pos=0.7,left,font=\ttfamily,blue!50!black] {\#pragma} (codegen);
		\path (codegen) edge[edge,dashed,bend right=80,draw=blue!50!black,postaction={decorate,decoration={text along path,text={|\color{blue!50!black}|Loop metadata},raise=-1.7ex,pre=moveto,pre length=13mm}}] (ir);
		\path (ir) edge[edge,dashed,draw=blue!50!black] (loopopts);
		\path (ir) edge[edge,dashed,draw=blue!50!black,bend right=10] (vectorization);
		\path (polly) edge[edge,dashed,draw=blue!50!black,bend right=10] (vectorization);
		\end{tikzpicture}%
	}
\vspace*{-1.5ex}%
\caption{Clang/LLVM compiler pipeline}
\label{fig:pipeline}
\vspace*{-2ex}%
\end{figure}

Clang/LLVM has a classical front-/mid- and back-end architecture as seen in \cref{fig:pipeline}. 
Clang, the front-end, translates the source code to LLVM-IR. 
In the mid-end, LLVM's optimization passes transform the IR itself.

OpenMP parallelism is processed in the front-end's IR generation. 
Clang outlines the parallel loop bodies/tasks into their own functions and passes them as function pointers to the OpenMP runtime. 
As an exception, \cinline{#pragma omp simd} annotates loops using \textinline{llvm.loop.vectorize.enable} metadata to be processed later by passes in the mid-end.
The \cinline{#pragma clang loop} directives also emit such metadata.

The currently available passes (Clang/LLVM 7.0) are: LoopUnswitch, LoopIdiom, LoopDeletion, LoopInterchange, SimpleLoopUnroll, VersioningLICM, LoopDistribute, LoopVectorize, LoopUnrollAndJam and LoopUnroll. 
The LoopVectorize pass looks for the \textinline{llvm.loop.vectorize.enable} metadata which overrides its cost model heuristic. 
Polly~\cite{polly} is a polyhedral loop optimizer capable of applying a plethora of loop transformations including most of those implemented by the dedicated passes but is not in the default pass pipeline and currently cannot be directed using pragmas.


Compiler writers like the concept of mostly independent optimization passes that can be arranged depending on the optimization level, but in case of loop transformations, this comes with several drawbacks.

\subsubsection{No optimization of OpenMP-parallel codes.}

Clang's IR generation of OpenMP emits function calls to runtime libraries.
This is because there is no construct in LLVM-IR to express thread-parallel execution.
The mid-end only sees that some function with opaque implementation is called, hence cannot change it. 

\noindent\textbf{Predefined transformation order.}
Transformations can only be applied in the order they appear in the pass pipeline, including the thread-parallelization via OpenMP as the first step in the front-end. 
For instance, LoopDistribute might separate a parallel loop from a sequential computation.
It is not possible to parallelize this loop since it must have had happened in the front-end.
It can still be vectorized because the LoopVectorize pass is executed after LoopDistribute.

\noindent\textbf{Order is implementation-defined.}
The pass order may depend on the optimization setting, but also change between versions of the compiler.
Generally, user source code should not depend on implementation-details such as the pass pipeline order.
One can only reliably apply a single transformation per loop.




%

\noindent\textbf{Separate infrastructure.}
The independence of the passes also has the disadvantage that passes do not reuse already obtained loop analysis results.
LLVM has a concept of analysis passes and analysis preservation, but it is not exploited, such that e.g. dependence analysis has to be recomputed for every loop pass.
LLVM currently has three different implementations of dependence analysis, each with different APIs.

\noindent\textbf{Code blowup through versioning.}
Loop passes invoke LoopVersioning themselves to get a clone of the instructions and control flow they want to transform.
That is, each loop transformation makes its own copy of the code, with a potential code growth that is exponential in the number of passes if the versioned fallback code is again optimized by further loop passes. 
In addition, many versioning conditions, such as checking for array aliasing, will be similar or equal for each versioning.

\noindent\textbf{Independent profitability heuristics.}
Another result of pass independence is that they only consider the profitability of a transformation ignoring transformation opportunities. 
For instance, loop distribution is likely only profitable if at least one of the successor loops is parallelizable or at least vectorizable, but the pass itself does not know whether a loop will be vectorized.

\noindent\textbf{IR normalization.}
LLVM's pass pipeline does not distinguish between IR normalization and optimization. 
Some optimizations such as Loop-Invariant Code Motion occur before the loop passes which make their work harder, as illustrated by \cref{lst:bakein}.

\begin{listing}
\begin{subfigure}[t]{.45\linewidth}
\begin{minted}{c}
for (int j=0; j<m; j+=1) {
  for (int i=0; i<n; i+=1) 
    A[i] += i*B[j];
}

\end{minted}
\vspace*{-2ex}%
\begin{tikzpicture}%
\node[draw=none] at (-1,0) {};
\path (0,0) edge[->,thick] node[right] {LICM} (0,-0.6);
\end{tikzpicture}%
\vspace*{-2ex}%
\begin{minted}{c}
for (int j=0; j<m; j+=1) {
  tmp = B[j];
  for (int i=0; i<n; i+=1) 
    A[i] += i*tmp;
}
\end{minted}
\end{subfigure}%
\begin{subfigure}[t]{.1\linewidth}%
\begin{tikzpicture}[overlay]%
\path (-0.75,0) edge[<->,thick] node[below] {\minibox[c]{Loop\\Interchange}} (0.75,0);
\end{tikzpicture}%
\end{subfigure}%
\begin{subfigure}[t]{.45\linewidth}%
\begin{minted}{c}
for (int i=0; i<n; i+=1) {
  for (int j=0; j<m; j+=1) 
    A[i] += i*B[j];
}

\end{minted}
\vspace*{-2ex}%
\begin{tikzpicture}%
\node[draw=none] at (-1,0) {};
\path (0,0) edge[->,thick] node[right] {Register Promotion} (0,-0.6);
\end{tikzpicture}%
\vspace*{-2ex}%
\begin{minted}{c}
for (int i=0; i<n; i+=1) {
  tmp = A[i];
  for (int j=0; j<m; j+=1) 
    tmp += i*B[j];
  A[i] = tmp;
}
\end{minted}
\end{subfigure}
\caption{Illustration of scalar passes that bake-in loop structures: The top codes can be converted to each other using loop interchange, but this is not directly possible after LICM or register promotion have been applied.}
\label{lst:bakein}
\end{listing}

\section{Loop Optimization Infrastructure}

To avoid the problems mentioned in the previous section, we propose a single pass dedicated to higher-level optimizations than the normal instruction- and control-flow based passes.
In its center is a new data structure we call \emph{Loop Structure DAG}. 
With it comes support functionality to traverse, analyze and modify it. 

\subsection{The Loop Structure DAG}

The Loop Structure DAG is based on LLVM's LoopInfo tree, an analysis for finding natural loops. 
Irreducible loops cannot be represented in the LoopInfo tree, but such loops can be converted to natural loop by copying basic blocks, similar to what LoopRotation already does.
Generally, loops must be assumed to have an infinite number of iterations, which can be limited by an exit conditions (i.e. while-loops) and/or a statically determinable number of executions, such as canonical for-loops.

In addition to loops, the DAG also contains statements and expressions. 
An \emph{expression} is any LLVM instruction that does not have side-effects, such as most arithmetic operations. 
Expressions do not have an execution time associated with it and can be materialized when necessary.

\begin{figure}[t]
\resizebox{\linewidth}{!}{%
\begin{tikzpicture}%
\tikzset{tight/.style={inner sep=0pt,outer sep=0pt,minimum size=0pt}}
\tikzset{ctrlnode/.style={font=\color{blue}}}
\tikzset{stmtnode/.style={font=\color{darkmagenta}}}
\tikzset{exprnode/.style={font=\color{darkgoldenrod}}}
\tikzset{edge/.style={->}}
\tikzset{crossedge/.style={crimsonglory}}

\graph[layered layout,edges={edge,rounded corners},level sep=5mm,sibling sep=10mm]{
	func [as={Function}];
	iloop [ctrlnode,as={\texttt{for (int i=0; i<128; i++)}}];
	jloop [ctrlnode,as={\texttt{for (int j=0; j<64; j++)}}];
	kloop [ctrlnode,as={\texttt{for (int k=0; k<256; k++)}}];
	Aassign [stmtnode,as={\texttt{A[i][j] = \dots}}];
	Bassign [stmtnode,as={\texttt{B[i][k] = \dots}}];
	jsin [exprnode,as={\texttt{j*sin(\dots)}}];
	kcos [exprnode,as={\texttt{k*cos(\dots)}}];
	pi [exprnode,as={\texttt{2*PI*i/128}}];
	func->iloop->{jloop,kloop};
	jloop->Aassign->jsin;
	kloop->Bassign->kcos;
	{jsin,kcos}->pi;
	funcp [as={Function'}];
	iloopp [ctrlnode,as={\texttt{for (int i=0; i<128; i++)}}];
	kloopp [ctrlnode,as={\texttt{for (int k=256; k>=0; i--)}}];
	funcp->iloopp->{kloopp};
	iloopp->[crossedge]jloop;
	kloopp->[crossedge]Bassign;
};
\end{tikzpicture}
}
\caption{A Loop Structure DAG; Control nodes are colored blue, statements in purple and expressions in yellow. 
The expression node \cinline{2*PI*i/128} is shared by two statement nodes. This is a form of Common Subexpression Elimination and IR code generator may try share the instructions for both uses. 
The modified Function \texttt{Function'} reuses some of the nodes of the original \texttt{Function}.
}
\label{fig:looptree}
\end{figure}
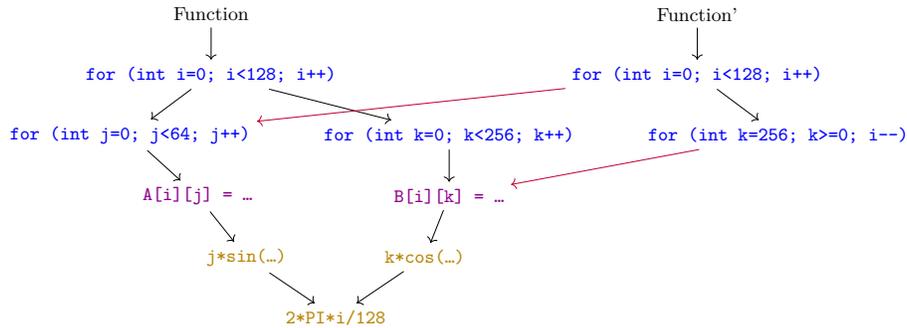

Consequently, instructions with side-effects are represented through \emph{statement} nodes. 
Statements and loops can have multiple children which represent their sequential execution. 
%
\Cref{fig:looptree} shown an illustration of the following Loop Structure DAG.
\vspace*{-1.5ex}%
\begin{minted}[]{c}
void Function() {
    for (int i = 0; i < 128; i+=1) {
        for (int j = 0; j <  64; j+=1) A[i][j] = j*sin(2*PI*i/128); 
        for (int k = 0; k < 256; k+=1) B[i][j] = j*cos(2*PI*i/128); 
    }
}
\end{minted}
\vspace*{-1.5ex}%

Nodes can be expanded lazily on-demand and are reusable: Copies of the DAG can be made by referencing the original DAG's nodes. 
Using copy-on-write semantics, changing a node in the DAG does not require copying the entire DAG, but only nodes on the path to the root. 
In the example of \cref{fig:looptree}, the for-loop over \texttt{k} has reversed which results in a modified DAG \texttt{Function'}. 

Reusable subtrees do not allow nodes to reference their parent nodes which would make a DAG traversal difficult. 
Fortunately, the red-green tree idea~\cite{redgreen} solves this issue, in our case a red-green DAG.
The green DAG is the tree with nodes having a list of all their children; that is, the DAG in \cref{fig:looptree}. 
The red DAG is a parallel tree where nodes reference a green node and the red node of its parent. 
It serves as a facade of the green DAG with additional parenthood information.


There are no control structures other than loops. 
Conditional execution is mapped to predicates: Every statement is annotated with a predicate that must evaluate to true or false defining whether the statement is executed.
The same predicates can be reused for multiple statements such that the IR generator can recover any acyclic control flow.
This means that expressions and control flow within a loop body is closer to a Sea-of-Nodes representation than LLVM IR.

Statement nodes have properties describing their behavior. 
The \emph{idempotent} property describes a statement that can be executed multiple times (with the same arguments), if known that it is executed at least once. 
LLVM instructions can also be \emph{speculatable}, which in the Loop Structure DAG would be represented as an expression.
An example for an idempotent, but not speculatable statement is division: If the argument executed at least once, we know that the execution did not trap, i.e. is non-zero, and hence can be executed again. 
Another example are memory loads and stores without dependencies. 
The property is useful for materializing the same value for different uses instead of storing the value to memory, which would create a dependency.

Loops can be marked as being executable in parallel, a quick way to determine these properties without looking at dependencies. 
The flag is pre-set if the Loop Structure DAG is created from am OpenMP-annotated loop.

If a node contains a construct that is hard to transform, it can be marked as such. 
For instance, a transformation might not want to handle exceptions and can skip nodes that are marked as exception-throwing. 
It is important that the IR can still be represented as a Loop Structure DAG such that unrelated code parts can still be transformed.


\subsection{Supporting Infrastructure}

Part of creation of a Loop Structure Graph is its normalization to a form with the most freedom for transformations, ideally such that all DAGs for \cref{lst:bakein} are equivalent.

A dependency analysis determines hazards between dependences as accurate as the optimization level allows -- from coarse-grained statement node dependencies to per-instance dependencies by polyhedral analysis~\cite{polly}.
In addition to memory dependencies, special types of dependencies allow better handling of common situations. 
Register dependencies track the data flow over virtual registers and phi-nodes.
If after transformations register lifetimes are overlapping, the IR generator can expand these to arrays.
Control dependencies can be emitted as either bit flags or branch instructions.

Thereafter, we collect properties about the loop nest, such as identifying common idioms like matrix-multiplication. 
This can trigger well-known optimizations or be replaced by hand-optimized library calls such as \texttt{gemm}.

The infrastructure should make it easy to implement many transformations.
These can either be triggered by source code directives such as suggested in~\cite{iwomp18-pragmas}, determined automatically using techniques from the polyhedral model~\cite{polly}, or selected by a hand-written function that captures the wisdom  beneficial loop transformations. 
For the latter it is essential that applying a transformation is safe and simple such that it can be tuned with the compiler's development without worrying about miscompilation and complexity.

At the end, the Loop Structure DAG is translated to either LLVM-IR or VPlan~\cite{vplan} (for vectorization). 
Multiple versions of the same code can be emitted, depending on which assumptions were made during analysis or transformation, and the cost model to select the best out of multiple DAGs.

\section{Related Work}

The Loop Structure DAG was inspired by other data structures: 
VPlan~\cite{vplan} is also designed to allow multiple instances (in this case for different vector widths) and estimate their profitability.
Isl's schedule trees~\cite{verdoolaege14}, which are also used by Polly, features a copy-on-write mechanism, but instead of red nodes, the tree is only accessibly through tree-iterators.
Silicon Graphics' compiler~\cite{sgi-lno} featured a Loop Nest optimizer component (LNO) with support for many pragma transformations.
IBM's xl high-order transformations are performed on a Loop Structure Graph~\cite{vivek97}, which follows the same idea as an alternative code representation, but without predicate-only control flow, red-green nodes or sea-of-nodes.

{\footnotesize
\subsubsection*{Acknowledgments.}

This research was supported by the Exascale Computing Project (17-SC-20-SC), a collaborative effort of two U.S. Department of Energy organizations (Office of Science and the National Nuclear Security Administration) responsible for the planning and preparation of a capable exascale ecosystem, including software, applications, hardware, advanced system engineering, and early testbed platforms, in support of the nation’s exascale computing imperative.

This research used resources of the Argonne Leadership Computing Facility, which is a DOE Office of Science User Facility supported under Contract DE-AC02-06CH11357.
}
